\documentclass{jpsj2}
\title{Full Self-Consistent Projection Operator Approach to Nonlocal
Excitations in Solids}

\author{
Yoshiro \textsc{Kakehashi}$^{1}$\thanks{E-mail address:
yok@sci.u-ryukyu.ac.jp}, Tetsuro \textsc{Nakamura}$^{1}$, 
and Peter \textsc{Fulde}$^{2,3}$
}

\inst{
$^{1}$Department of Physics and Earth Sciences,
Faculty of Science, University of the Ryukyus, \\
1 Senbaru, Nishihara, Okinawa, 903-0213, Japan \\
$^{2}$Max-Planck-Institut f\"ur Physik komplexer Systeme,
N\"othnitzer Str. 38, D-01187 Dresden, Germany
$^{3}$Asia Pacific Center for Theoretical Physics,
Pohang, Gyeongbuk 790-784, Korea
}

\abst{
A self-consistent projection operator method for single-particle
excitations is developed.  It describes the nonlocal correlations
on the basis of a projection technique to the retarded Green
function and the off-diagonal effective medium.  The theory takes into
account long-range intersite correlations making use of an incremental
cluster expansion in the medium.  A generalized self-consistent coherent
potential is derived.  It yields the momentum-dependent
excitation spectra with high resolution.  Numerical studies for the Hubbard
model on a simple cubic lattice at half filling show that the theory
is applicable in a wide range of Coulomb interaction strength.  
In particular, it is found that the long-range antiferromagnetic
correlations in the strong interaction regime cause shadow bands
in the low-energy region and sub-peaks of the
Mott-Hubbard bands.
}

\kword{non-local excitations, momentum-dependent self-energy, Hubbard
model, ARPES}

\begin{document}
\maketitle

\section{Introduction} 
Single-particle excitations play an important role in condensed matter
physics.  They determine among others basic properties of solids such as
the metal-insulator (MI) transition, magnetism, and 
superconductivity~\cite{fulde}.
Recently developed angle-resolved photoemission
spectroscopy allows to observe details of the excitation
spectra in various materials~\cite{shen,imada}.
The excitations are usually strongly influenced by electron
correlations.  Therefore various approaches to treat the correlations 
have been developed.
Hubbard~\cite{hub63,hubbard3}, 
for example, was the first who proposed a theory of the MI
transition on the basis of the retarded Green function. He derived
lower and upper Mott-Hubbard incoherent bands caused by strong electron
correlations.  
Cyrot~\cite{cyrot} extended the theory to finite
temperatures by using the functional integral method.  
Penn~\cite{penn79} and Liebsch~\cite{lieb79} 
developed a Green function theory starting from the low-density 
limit using the t-matrix approximation.
Fulde {\it et al.}~\cite{luk82,becker90,unger93,unger94,fulde02} 
proposed methods which use the projection technique.  
The latter describes the dynamics of electrons by means of
the Liouville operator on the basis of the retarded Green functions.

In the past two decades single-site theories of excitations with use of
an effective medium have extensively been developed.  
Progress was made for the MI transition in infinite dimensions where 
the self-energy of the Green function becomes independent of 
momentum~\cite{metzner}.
Several authors~\cite{muller,jarrell,georges93,georges96} 
determined the self-energy ({\it i.e.}, an effective medium) 
self-consistently so as to be identical with the local
self-energy of an impurity embedded in a medium.  
The theory, called dynamical mean-field theory (DMFT), 
can be traced back to a many-body 
coherent potential approximation (CPA) in disordered 
alloys~\cite{hirooka} and
is equivalent to the dynamical CPA~\cite{kake92,kake02} used 
in the theory of magnetism~\cite{kake022}.
The DMFT has clarified the MI transition in infinite dimensions.  
One of its new features is that it can describe both 
quasiparticle states near the Fermi level and the Mott-Hubbard
incoherent bands.  

In the DMFT as well as the dynamical CPA one usually deals with 
the temperature Green function.  
Therefore one needs to perform numerical analytic continuations at
finite temperatures, which often causes numerical difficulties 
in particular at low temperatures.  We recently proposed an 
alternative method which directly starts from the retarded Green function.
The method, which is called the projection operator method CPA 
(PM-CPA)~\cite{kake04-1},  
has been shown to be equivalent to the dynamical CPA and 
the DMFT~\cite{kake04-2}.  
In the PM-CPA, the Loiuville operator is approximated by an
energy-dependent Liouvillean for an effective Hamiltonian with a coherent
potential.  The latter is determined by a self-consistent CPA condition.
By solving an impurity problem embedded in a medium with use
of the renormalized perturbation theory (RPT), we obtained an
interpolation theory for the MI transition.

Although one can treat the excitations from the weak to the strong
Coulomb interaction regime with use of the single-site theories
mentioned above, the single-site approximation (SSA) 
neglects the intersite correlations which often play
an important role in real systems such as the Cu oxcides and 
Fe-pnictides high-temperature superconducting compounds.  
In fact, recent photoelectron spectroscopy found 
a pseudo gap~\cite{mars96,yoshida03} 
and a kink structure~\cite{bodanov,lanzara} 
in cuprates which can not be explained by a SSA.

Because of the reasons mentioned above, we have recently proposed a
nonlocal theory of excitation spectra called the self-consistent
projection operator method (SCPM)~\cite{kake04-3}. 
It is based on the projection technique~\cite{mori65,zwa61} 
and the incremental cluster expansion 
method~\cite{stoll90,grafen93,grafen97}.  
In this theory, all the off-diagonal self-energy 
matrix elements are calculated by means of
an incremental cluster expansion from a diagonal effective medium
$\tilde{\Sigma}_{ii\sigma}(z)$.  The latter is determined by the
CPA condition.  We call this method here and in the following 
the SCPM-0, The theory takes into account long-range intersite 
correlations, which are missing in the other 
nonlocal theories such as the dynamical cluster 
theory~\cite{hettler98,jarrell01} and the 
cluster DMFT~\cite{kotliar01,sun02}.  
Moreover, the theory can describe the momentum dependent excitation 
spectrum with high resolution because it is based on the retarded 
Green function and because the intersite correlations are taken into
account up to infinity at each order of the incremental cluster 
expansion.
Using the SCPM-0, we investigated the excitation spectra
of the Hubbard model on the simple cubic lattice~\cite{kake04-3} 
and the square lattice~\cite{kake05,kake05-2,kake07}.
Especially in the two-dimensional system, we found
a marginal Fermi liquid behavior~\cite{kake05} 
and a kink structure~\cite{kake05-2,kake07} 
in the quasiparticle state in the underdoped region. 
The former was proposed
in a phenomenological theory~\cite{varma89} and the latter was found 
in the photoemission experiments in cuprates~\cite{bodanov,lanzara}.

The SCPM-0 makes use of a diagonal effective medium to describe 
the on-site
correlations at surrounding sites.  The self-consistency between the 
self-energy and the medium is achieved only for the diagonal site.  
Such a treatment results in a limited range of applications.  
In this paper, we extend the SCPM-0 by introducing a new
off-diagonal effective medium $\tilde{\Sigma}_{ij\sigma}(z)$. 
All calculated off-diagonal self-energy matrix elements are consistent 
with all those of the medium.  
Such a full self-consistent projection operator method (FSCPM) should 
allow for an improved quantitative description of nonlocal excitations 
in solids, and should play an important role in the
phenomena with long-range charge and spin fluctuations,
in particular, in low dimensional systems.

In the following subsection 2.1, we briefly review the retarded Green
function and projection technique.  In the \S 2.2, we introduce an
energy-dependent Liouville operator whose corresponding 
Hamiltonian consists of the
noninteracting Hamiltonian and an off-diagonal effective medium.   
We then expand the nonlocal self-energy embedded in the off-diagonal 
effective medium using the incremental cluster method.
In order to obtain the different terms of the expansion,
we need to calculate cluster memory functions for the medium.  
In \S 2.3 we obtain them by making use of the renormalized perturbation 
scheme.  The off-diagonal medium is determined by a fully self-consistent
condition. 
We present in \S 3 numerical results for momentum dependent excitation
spectra for the Hubbard model on a simple cubic lattice at half-filling,
and demonstrate that the full self-consistency extends the range of
application from the intermediate Coulomb interaction regime 
to the strongly correlated regime.
Calculated spectra modify the previous results based on the SCPM-0 
so as to suppress the nonlocal effects in the intermediate regime.
In the strong Coulomb interaction regime, we find the shadow bands
associated with strong antiferromagnetic (AF) correlations and 
the band splitting in the Mott-Hubbard incoherent bands.  
In the last section we
summalize the numerical results and discuss future problems.

\section{Full Self-Consistent Projection Operator Method}
\subsection{Retarded Green function and projection technique}
We adopt the tight-binding Hubbard model consisting of the Hartree-Fock
independent particle Hamiltonian $H_{0}$ and the residual interactions
with intraatomic Coulomb interaction parameter $U$ as follows:
\begin{eqnarray}
H = H_{0} + U \sum_{i} \, \delta n_{i \uparrow} \delta n_{i \downarrow} \ ,
\label{hub}
\end{eqnarray}
\begin{eqnarray}
H_{0} = 
\sum_{i,\sigma} \, \epsilon_{\sigma} \, n_{i \sigma}
+ \sum_{i, j, \sigma} t_{i j} \, a_{i \sigma}^{\dagger} 
a_{j \sigma} \ .
\label{hub-0}
\end{eqnarray}
Here 
$\epsilon_{\sigma} = \epsilon_{0} - \mu + U \langle n_{i-\sigma} \rangle$ 
is the Hartree-Fock atomic level. 
Note that $\langle \cdots \rangle$ denotes a thermal average. 
The quantities 
$\epsilon_{0}$, $\mu$, and $t_{i j}$ are the atomic level, the chemical
potential, and the transfer integral between sites $i$ and $j$, 
respectively.  Furthermore
$a_{i \sigma}^{\dagger}$ ($a_{i \sigma}$)
denotes the creation (annihilation) operator for an electron with spin
$\sigma$ on site $i$, while  
$n_{i \sigma}=a_{i \sigma}^{\dagger}a_{i \sigma}$ is the
electron density operator for spin $\sigma$, and 
$\delta n_{i \sigma} = n_{i \sigma} - \langle n_{i \sigma} \rangle$.

The excitation spectra for electrons are obtained from a retarded Green
function defined by
\begin{eqnarray}
G_{ij\sigma}(t-t^{\prime}) = 
-i \theta(t-t^{\prime}) 
\langle [ a_{i\sigma}(t), a^{\dagger}_{j\sigma}(t^{\prime})  ]_{+}
\rangle \ .
\label{rg1}
\end{eqnarray}
Here $\theta(t)$ is the step function, $a_{i\sigma}(t)$ is the Heisenberg
representation of $a_{i\sigma}$ defined by 
$a_{i\sigma}(t)= \exp(iHt) a_{i\sigma} \exp(-iHt)$.  Furthermore, 
$[\ , \ ]_{+}$ denotes the anti-commutator between the operators.

The Fourier representation of the retarded Green function is written as
follows~\cite{fulde}. 
\begin{eqnarray}
G_{ij\sigma}(z) = 
\Big( a^{\dagger}_{i \sigma} \  {\Big |} \ \frac{1}{z-L} \,
a^{\dagger}_{j \sigma} \Big) \ .
\label{rg2}
\end{eqnarray}
Here the inner product between the operators $A$ and $B$ is defined by 
$(A|B)=\langle [A^{+},B]_{+} \rangle$.  Also 
$z=\omega+i\delta$ with $\delta$ being an infinitesimal positive number,
and $L$ is a Liouville operator defined by $LA=[H,A]$ for an operator
$A$.  Note that $[\ , \ ]$ is the commutator between the operators.

Using the projection technique, we obtain the Dyson equation
for the retarded Green function as
\begin{eqnarray}
G_{ij\sigma}(z) = [(z - \mbox{\boldmath$H$}_{0} 
- \mbox{\boldmath$\Lambda$}(z))^{-1}]_{ij\sigma} \ .
\label{eqmt1}
\end{eqnarray}
Here the matrices are defined by 
$(\mbox{\boldmath$H$}_{0})_{ij\sigma} = 
\epsilon_{\sigma} \, \delta_{ij} + t_{i j}$, and
\begin{eqnarray}
\Lambda_{ij\sigma}(z) = (\mbox{\boldmath$\Lambda$}(z))_{ij\sigma} = 
U^{2} \overline{G}_{ij\sigma}(z) \ . 
\label{self}
\end{eqnarray}
The reduced memory function $\overline{G}_{ij\sigma}(z)$ is defined by
\begin{eqnarray}
\overline{G}_{ij\sigma}(z) = 
\Big( A^{\dagger}_{i \sigma} \  {\Big |} \ \frac{1}{z-\overline{L}} \,
A^{\dagger}_{j \sigma} \Big) \ .
\label{rmem}
\end{eqnarray}
The operator $A^{\dagger}_{i \sigma}$ is given by 
\begin{eqnarray}
A^{\dagger}_{i \sigma}=a^{\dagger}_{i \sigma} \delta n_{i -\sigma} \ ,
\label{ai}
\end{eqnarray}
and the Liouville operator $\overline{L}$ 
acting on $A^{\dagger}_{j \sigma}$ is defined
by $\overline{L}=QLQ$, $Q$ being $Q=1-P$.
The projection operator $P$ projects onto the original operator
space $\{ |a^{\dagger}_{i \sigma}) \}$,
\begin{eqnarray}
P= \sum_{i\sigma} 
\big| a^{\dagger}_{i \sigma} \big) \, \big(a^{\dagger}_{i \sigma} \big|
\ .
\label{p1}
\end{eqnarray}

In a crystalline system, the Green function $G_{ij\sigma}(z)$ is
obtained from its momentum representation $G_{k\sigma}(z)$ as 
\begin{eqnarray}
G_{ij\sigma}(z) = \frac{1}{N} \sum_{k} 
G_{k\sigma}(z) 
\exp (i\mbox{\boldmath$k$}\cdot(\mbox{\boldmath$R$}_{i}
-\mbox{\boldmath$R$}_{j})) \ .
\label{gf}
\end{eqnarray}
Here $N$ is a number of site.
The momentum-dependent Green function is given by 
\begin{eqnarray}
G_{k\sigma}(z) = \frac{1}{z - \epsilon_{k\sigma} - 
\Lambda_{k\sigma}(z)} \ .
\label{gk}
\end{eqnarray}
Here $\epsilon_{k\sigma}=\epsilon_{\sigma} + \epsilon_{k}$ 
is the Hartree-Fock one-electron energy eigenvalue, 
$\epsilon_{k}$ is the Fourier transform of $t_{ij}$,
and 
\begin{eqnarray}
\Lambda_{k\sigma}(z) = \sum_{j} \Lambda_{j0\sigma}(z) 
\exp (i\mbox{\boldmath$k$}\cdot\mbox{\boldmath$R$}_{j}) \ .
\label{lk}
\end{eqnarray}

It should be noted that the momentum-dependent excitation spectrum is
obtained from
\begin{eqnarray}
\rho_{k\sigma}(\omega) = -\frac{1}{\pi}\, {\rm Im}\, G_{k\sigma}(z) \ .
\label{rhokk}
\end{eqnarray}
The local density of states (DOS) is given by
\begin{eqnarray}
\rho_{i\sigma}(\omega) = -\frac{1}{\pi}\, {\rm Im}\, G_{ii\sigma}(z) \ .
\label{rhoii}
\end{eqnarray}
which is identical with the average DOS per atom, 
$\rho_{\sigma}(\omega) = 1/N \sum_{k} \rho_{k\sigma}(\omega)$
when all sites are equivalent to each other.

\subsection{Incremental cluster expansion in the off-diagonal effective
  medium} 

In the full self-consistent projection method, we introduce an
energy-dependent Liouville operator $\tilde{L}(z)$.  Its corresponding
Hamiltonian is that of an off-diagonal
effective medium $\tilde{\Sigma}_{ij\sigma}(z)$.  It is 
\begin{eqnarray}
\tilde{L}(z) A = [\tilde{H}_{0}(z),A] \ ,
\label{leff}
\end{eqnarray}
for arbitrary operator $A$ and 
\begin{eqnarray}
\tilde{H}_{0}(z) = H_{0} + \sum_{ij\sigma} 
\tilde{\Sigma}_{ij\sigma}(z) \, a^{\dagger}_{i \sigma}a_{j \sigma} \ .
\label{heff}
\end{eqnarray}
Defining an interaction Liouville operator $L_{\rm I}(z)$ such that 
\begin{eqnarray}
L_{\rm I}(z) A = [H_{\rm I}(z),A] \ ,
\label{li0}
\end{eqnarray}
with
\begin{eqnarray}
H_{\rm I}(z) = U \sum_{i} \delta n_{i\uparrow} \delta n_{i\downarrow}
- \sum_{ij\sigma} \tilde{\Sigma}_{ij\sigma}(z) 
a^{\dagger}_{i\sigma} a_{j\sigma} \ ,
\label{hieff}
\end{eqnarray}
we can rewrite the original Liouville operator $L$ as follows.
\begin{eqnarray}
L = \tilde{L}(z) + L_{\rm I}(z) \ . 
\label{lsep}
\end{eqnarray}

It should be noted that the interaction $H_{\rm I}(z)$ contains the
off-diagonal components $\tilde{\Sigma}_{ij\sigma}(z) \ (i \neq j)$ 
in addition to the diagonal ones $\tilde{\Sigma}_{ii\sigma}(z)$.  
Accordingly, we divide here the interaction Liouvillean
$L_{\rm I}(z)$ into single-site terms and pair-site ones.
Furthermore we 
introduce site-dependent prefactors $\{ \nu_{i} \}$ which are 
either 1 or 0.
\begin{eqnarray}
L_{\rm I}(z) = \sum_{i} \nu_{i} \ {\mathcal L}^{(i)}_{\rm I}(z) + 
\sum_{(i,j)} \nu_{i}\nu_{j} \ {\mathcal L}^{(ij)}_{\rm I}(z) \ , 
\label{lintsep}
\end{eqnarray}
\begin{eqnarray}
{\mathcal L}^{(i)}_{I}(z) A = \Bigl[
U \delta n_{i\uparrow} \delta n_{i\downarrow} -
\sum_{\sigma} \tilde{\Sigma}_{ii\sigma}(z) n_{i\sigma} \, , A \Bigr] \ ,
\label{tli}
\end{eqnarray}
\begin{eqnarray}
{\mathcal L}^{(ij)}_{I}(z) A = \Bigl[ -
\sum_{\sigma} (\tilde{\Sigma}_{ij\sigma}(z) 
a^{\dagger}_{i\sigma}a_{j\sigma} + \tilde{\Sigma}_{ji\sigma}(z) 
a^{\dagger}_{j\sigma}a_{i\sigma}) \, , A \Bigr] \ .
\label{lij}
\end{eqnarray}

After substituting the Liouvillean (\ref{lsep}) into
Eq. (\ref{rmem}), we can expand the resolvent $(z-\overline{L})^{-1}$ 
with respect to $L_{\rm I}(z)$ as follows.
\begin{eqnarray}
(z-\overline{L})^{-1} = \overline{G}_{0} 
+ \overline{G}_{0}T\overline{G}_{0} \ ,
\label{lexp}
\end{eqnarray}
\begin{eqnarray}
T = \overline{L}_{I} 
+ \overline{L}_{I}\overline{G}_{0}\overline{L}_{I} 
+ \overline{L}_{I}\overline{G}_{0}\overline{L}_{I}\overline{G}_{0}
\overline{L}_{I} + \cdots \ . 
\label{texp1}
\end{eqnarray}
Here $\overline{G}_{0}=(z-\overline{L}_{0}(z))^{-1}$ and 
$\overline{L}_{0}(z)=Q\tilde{L}(z)Q$, while 
$\overline{L}_{I}(z)=QL_{\rm I}(z)Q$.

The $T$ matrix operator may be expanded with respect to different sites as
\begin{eqnarray}
T = \sum_{i} \nu_{i}T_{i} 
+ \sum_{(i,j)} \nu_{i}\nu_{j}T_{ij}
+ \sum_{(i,j,k)} \nu_{i}\nu_{j}\nu_{k}T_{ijk} + \cdots \ .
\label{texp2}
\end{eqnarray}
The single-site $T_{i}$, two-site $T_{ij}$, and three-site $T_{ijk}$ 
matrix scattering operators are obtained by setting the indices as 
$(\nu_{i}=1, \nu_{l (\ne i)}=0)$, 
$(\nu_{i}=\nu_{j}=1, \nu_{l (\ne i,j)}=0)$, 
$(\nu_{i}=\nu_{j}=\nu_{k}=1, \nu_{l (\ne i,j,k)}=0)$, and so on.
It is 
\begin{eqnarray}
T_{i} = T^{(i)} \ ,
\label{ti}
\end{eqnarray}
\begin{eqnarray}
T_{ij} = T^{(ij)} - T_{i} - T_{j} \ ,
\label{tij}
\end{eqnarray}
\begin{eqnarray}
T_{ijk} = T^{(ijk)} - T_{ij} - T_{jk} - T_{ki} - T_{i} -T_{j} - T_{k} \ .
\label{tijk}
\end{eqnarray}
The operator $T^{({\rm c})}$ $({\rm c} = i,ij,ijk, \cdots)$ at the
right-hand-side (r.h.s.) of the above equations is the $T$ matrix 
operators for the cluster c, {\it i.e.}, 
\begin{eqnarray}
T^{({\rm c})} = 
\overline{L}^{({\rm c})}_{I}
(1-\overline{G}_{0}\overline{L}^{({\rm c})}_{I})^{-1} \ .
\label{tc}
\end{eqnarray}
Here $\overline{L}^{({\rm c})}_{I}=QL^{({\rm c})}_{\rm I}(z)Q$ and 
$L^{({\rm c})}_{\rm I}(z)$ is the interaction Liouvillean for 
a cluster c:
\begin{eqnarray}
L^{({\rm c})}_{\rm I}(z) = 
\sum_{i \in {\rm c}} {\mathcal L}^{(i)}_{\rm I}(z) + 
\sum_{(i,j) \in {\rm c}} {\mathcal L}^{(ij)}_{\rm I}(z) \ . 
\label{lic}
\end{eqnarray}
Note that the sums in the above equation are taken over the sites or 
pairs belonging to the cluster c.

Substituting Eq. (\ref{texp2}) into Eq. (\ref{lexp}) we have 
\begin{eqnarray}
\overline{G}_{ij\sigma}(z, \{ \nu_{l}\}) =  
\Big( A^{\dagger}_{i \sigma} \  {\Big |} 
(\overline{G}_{0} 
+ \sum_{l}\nu_{l}\overline{G}_{0}T_{l}\overline{G}_{0}
+ \sum_{(l,m)}\nu_{l}\nu_{m}\overline{G}_{0}T_{lm}\overline{G}_{0}
+ \cdots )
A^{\dagger}_{j \sigma} \Big) \ .
\label{gnu}
\end{eqnarray}
In the incremental method~\cite{stoll90,grafen93,grafen97}, 
we first consider the self-energy
contribution due to intra-atomic excitations, {\it i.e.},
\begin{eqnarray}
\overline{G}^{(i)}_{ii\sigma}(z) = 
\overline{G}_{ii\sigma}(z, \nu_{i}=1, \nu_{l(\neq i)}=0) =  
\big( A^{\dagger}_{i \sigma} \  {\big |} 
(\overline{G}_{0} 
+ \overline{G}_{0}T_{i}\overline{G}_{0})
A^{\dagger}_{i \sigma} \big) \ .
\label{gintra}
\end{eqnarray}
Next we consider the scattering contribution due to a two-site
increment in Eq. (\ref{gnu}),
\begin{eqnarray}
\overline{G}^{(il)}_{ii\sigma}(z) = 
\overline{G}_{ii\sigma}(z, \nu_{i}=\nu_{l}=1, \nu_{m(\neq i,l)}=0) = 
\overline{G}^{(i)}_{ii\sigma}(z)
+ \big( A^{\dagger}_{i \sigma} \  {\big |} 
\overline{G}_{0}(T_{l} + T_{li})\overline{G}_{0}
{\big |} A^{\dagger}_{i \sigma} \big) \ .
\label{gpii}
\end{eqnarray}
This defines two-site increment to the diagonal matrix element as
\begin{eqnarray}
\Delta\overline{G}^{(il)}_{ii\sigma}(z) =
\overline{G}^{(il)}_{ii\sigma}(z) - \overline{G}^{(i)}_{ii\sigma}(z) \ .
\label{icr2ii}
\end{eqnarray}

In the same way, we consider
\begin{eqnarray}
\overline{G}^{(ilm)}_{ii\sigma}(z) & = &
\overline{G}^{(i)}_{ii\sigma}(z)
+ \Delta\overline{G}^{(il)}_{ii\sigma}(z)
+ \Delta\overline{G}^{(im)}_{ii\sigma}(z)
+ \big( A^{\dagger}_{i \sigma} \  {\big |} 
\overline{G}_{0}(T_{lm} + T_{ilm})\overline{G}_{0}
{\big |} A^{\dagger}_{i \sigma} \big) \ .
\label{g3ii}
\end{eqnarray}
Then, we define the increment for a three-site contribution.
\begin{eqnarray}
\Delta\overline{G}^{(ilm)}_{ii\sigma}(z) & = & 
\overline{G}^{(ilm)}_{ii\sigma}(z) 
- \Delta\overline{G}^{(il)}_{ii\sigma}(z) 
- \Delta\overline{G}^{(im)}_{ii\sigma}(z) 
- \overline{G}^{(i)}_{ii\sigma}(z)\ .
\label{icr3ii}
\end{eqnarray}
The memory function $\overline{G}_{ii\sigma}(z)$ in Eq. (\ref{gnu}) is 
then expanded as follows.
\begin{eqnarray}
\overline{G}_{ii \sigma}(z) &=&
\overline{G}^{(i)}_{ii \sigma}(z) + 
\sum_{l \neq i} \Delta \overline{G}^{(il)}_{ii \sigma}(z)
+ \, \frac{1}{2} \, {\sum_{l \neq i}} \,{\sum_{m \neq i,l}} \, 
\Delta \overline{G}^{(ilm)}_{ii \sigma}(z) + \cdots \ .
\label{icrii}
\end{eqnarray}
It should be noted that $\overline{G}^{(\rm c)}_{ii\sigma}(z)$ 
$({\rm c}= i, ij, \cdots)$ in Eqs. (\ref{gintra}), (\ref{gpii}), and 
(\ref{g3ii}) is obtained from the $T$-matrix $T^{(\rm c)}$ given by
Eq. (\ref{tc}) as
\begin{eqnarray}
\overline{G}^{({\rm c})}_{ij\sigma}(z) =  
\big( A^{\dagger}_{i\sigma} \  {\big |} 
(z-\overline{L}^{({\rm c})}(z))^{-1} \,
A^{\dagger}_{j\sigma} \big) \ .
\label{gcij}
\end{eqnarray}
Here the cluster Liouvillean $\overline{L}^{({\rm c})}(z)$ is defined by
\begin{eqnarray}
\overline{L}^{({\rm c})}(z) = \overline{L}_{0}(z) 
+ \overline{L}^{({\rm c})}_{\rm I}(z) \ . 
\label{lc}
\end{eqnarray}

In the same way, the off-diagonal memory function is obtained as
follows.
\begin{eqnarray}
\overline{G}_{ij \sigma}(z) &=&
\overline{G}^{(ij)}_{ij \sigma}(z) + 
\sum_{l \neq i,j} \Delta \overline{G}^{(ijl)}_{ij \sigma}(z)
+ \, \frac{1}{2} \, \sum_{l \neq i,j} \sum_{m \neq i,j,l} \, 
\Delta \overline{G}^{(ijlm)}_{ij \sigma}(z) + \cdots \ ,
\label{icrij}
\end{eqnarray}
with
\begin{eqnarray}
\Delta\overline{G}^{(ijl)}_{ij\sigma}(z) = 
\overline{G}^{(ijl)}_{ij\sigma}(z) - \overline{G}^{(ij)}_{ij\sigma}(z) \ ,
\label{icr2ij}
\end{eqnarray}
\begin{eqnarray}
\Delta\overline{G}^{(ijlm)}_{ij\sigma}(z) = 
\overline{G}^{(ijlm)}_{ij\sigma}(z) 
- \Delta\overline{G}^{(ijl)}_{ij\sigma}(z) 
- \Delta\overline{G}^{(ijm)}_{ij\sigma}(z) 
- \overline{G}^{(ij)}_{ij\sigma}(z)\ .
\label{icr3ij}
\end{eqnarray}

When we take into account all the terms on the r.h.s. of
Eqs. (\ref{icrii}) and (\ref{icrij}), the memory function 
$\overline{G}_{ij \sigma}(z)$ does not depend on the effective medium 
$\tilde{\Sigma}_{ij\sigma}(z)$.  However, it is not possible in general to
calculate the terms up to higher orders; we have to truncate the 
incremental expansion at a certain stage.  In that case the memory function  
$\overline{G}_{ij \sigma}(z)$ depends on the medium 
$\tilde{\Sigma}_{ij\sigma}(z)$.  We determine the latter from the
following self-consistent equation,
\begin{eqnarray}
\tilde{\Sigma}_{ij\sigma}(z) = \Lambda_{ij\sigma}(z) \ .
\label{sceqij}
\end{eqnarray}
Note that the off-diagonal effective medium 
$\tilde{\Sigma}_{ij\sigma}(z)$ is the self-energy for the
energy-dependent Liouvillean $\tilde{L}(z)$, {\it i.e.},
$\tilde{\Sigma}_{ij\sigma}(z) = U^{2}  
\big( A^{\dagger}_{i\sigma} \  {\big |} 
(z-\overline{L}_{0}(z))^{-1} \,
A^{\dagger}_{j\sigma} \big) = U^{2}  
\big( A^{\dagger}_{i\sigma} \  {\big |} 
\overline{G}_{0} A^{\dagger}_{j\sigma} \big)$.
Thus the self-consistent equation (\ref{sceqij}) is equivalent to the
condition that the $T$-matrix describing the scattering from the
medium vanishes according to Eq. (\ref{lexp}):
\begin{eqnarray}
\big( A^{\dagger}_{i\sigma} \  {\big |} 
\overline{G}_{0} T \overline{G}_{0} A^{\dagger}_{j\sigma} \big) = 0 \ .
\label{gcpa}
\end{eqnarray}
This is a generalization of the CPA.

The present theory reduces to the previous version
of the nonlocal excitations (SCPM-0)~\cite{kake04-3} 
when the off-diagonal media 
$\tilde{\Sigma}_{ij\sigma}(z) \ (i \neq j)$ are omitted 
and only the self-consistency of
the diagonal part is taken into account in Eq. (\ref{sceqij}).
When in the SCPM-0 only the diagonal self-energy $\Lambda_{ii\sigma}(z)$ 
is taken into account ({\it i.e.}, when we make the SSA), the result 
reduces to the PM-CPA which we previously proposed~\cite{kake04-1}.

\subsection{Renormalized perturbation scheme to cluster memory functions}
The incremental cluster expansion scheme given 
in the last subsection
can be performed when the cluster memory function 
$\overline{G}^{({\rm c})}_{ij \sigma}(z)$ defined by
Eq. (\ref{gcij}) is known.  We obtain here an explicit expression for 
it.
The Hamiltonian $H^{({\rm c})}(z)$ to the cluster Liouvillean 
$L^{({\rm c})}(z)$ in (\ref{gcij}) is given by
\begin{eqnarray}
H^{\rm (c)} = H_{0} 
+ \sum_{ij\sigma} \tilde{\Sigma}_{ij\sigma}(z) 
a^{\dagger}_{i\sigma} a_{j\sigma} 
- \sum_{ij \in {\rm c}} \sum_{\sigma} \tilde{\Sigma}_{ij\sigma}(z) 
a^{\dagger}_{i\sigma} a_{j\sigma} + 
U \sum_{i \in {\rm c}} \, \delta n_{i \uparrow} \delta n_{i \downarrow} 
\ .
\label{hubc0}
\end{eqnarray}

\begin{figure}[tb]
\begin{center}
\includegraphics[width=14cm]{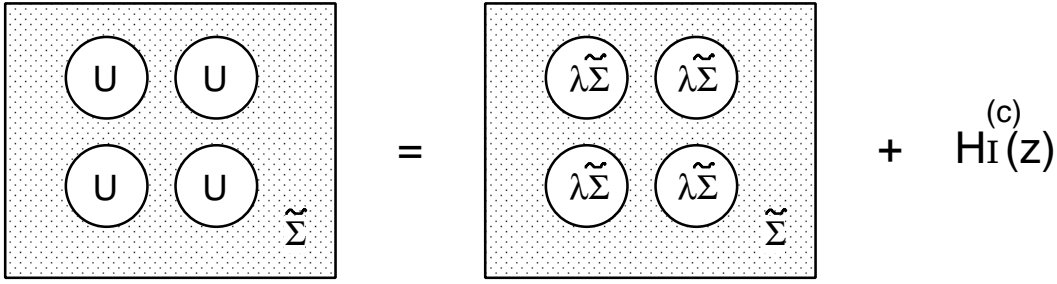}%
\end{center}
\caption{\label{figlmed}
Partition of the medium on the cluster sites.  The cluster Hamiltonian 
$H^{({\rm c})}(z)$ with on-site Coulomb interaction embedded in the
off-diagonal medium $\tilde{\Sigma}$ (the left-hand-side) can be split 
into an effective Hamiltonian $\tilde{H}^{({\rm c})}(z)$ with cluster
 potential $\lambda \tilde{\Sigma}$ and the remaining interaction 
$H^{({\rm c})}_{\rm I}(z)$.  The parameter $\lambda$ can control the
 magnitude of the potential $\tilde{\Sigma}$ on the cluster sites.
Note that each site in the cluster may be away from the other sites of 
the cluster. 
}
\end{figure}

Introducing parameters $\lambda_{ij\sigma}$ 
$(0 \le \lambda_{ij\sigma} \le 1)$, we can divide the Hamiltonian
$H^{(\rm c)}$ as 
\begin{eqnarray}
H^{({\rm c})}(z) = \tilde{H}^{({\rm c})}(z) + H^{({\rm c})}_{\rm I}(z) \ ,
\label{hubc}
\end{eqnarray}
\begin{eqnarray}
\tilde{H}^{({\rm c})}(z) = H_{0}
+ \sum_{ij\sigma} \tilde{\Sigma}_{ij\sigma}(z) 
a^{\dagger}_{i\sigma} a_{j\sigma} 
- \sum_{ij \in {\rm c}} \sum_{\sigma} \overline{\lambda}_{ij\sigma}
\tilde{\Sigma}_{ij\sigma}(z) 
a^{\dagger}_{i\sigma} a_{j\sigma} \ ,
\label{tihc0}
\end{eqnarray}
\begin{eqnarray}
H^{({\rm c})}_{\rm I}(z) = 
U \sum_{i \in {\rm c}} \, \delta n_{i \uparrow} \delta n_{i \downarrow} 
- \sum_{ij \in {\rm c}} \sum_{\sigma} \lambda_{ij\sigma}
\tilde{\Sigma}_{ij\sigma}(z) 
a^{\dagger}_{i\sigma} a_{j\sigma} \ .
\label{hci}
\end{eqnarray}
Here $\overline{\lambda}_{ij\sigma} = 1 - \lambda_{ij\sigma}$.
Note that parameters $\lambda_{ij\sigma}$ control the partition ratio of 
the medium potentials $\tilde{\Sigma}_{ij\sigma}$ in the cluster between
the noninteracting part 
$\tilde{H}^{({\rm c})}(z)$ and the interacting part 
$H^{({\rm c})}_{\rm I}(z)$ (see Fig. \ref{figlmed}). 
The Hamiltonian $\tilde{H}^{({\rm c})}(z)$ denotes a system with a
uniform effective medium $\tilde{\Sigma}_{ij\sigma}(z)$ when 
$\lambda_{ij\sigma} = 1 \ (i,j \in {\rm c})$.
When $\lambda_{ij\sigma} = 0 \ (i,j \in {\rm c})$, 
$\tilde{H}^{({\rm c})}(z)$ denotes a reference system with 
a cluster cavity in the effective medium and 
$H^{({\rm c})}_{\rm I}(z)$ denotes the Coulomb interactions on the cluster
sites.
%
%

According to the partition of the Hamiltonian (\ref{hubc}), we
introduce a {\it noninteracting} Liouvillean 
$\overline{L}^{({\rm c})}_{0}(z) = QL^{({\rm c})}_{0}(z)Q$ and 
an {\it interacting} Liouvillean 
$\overline{L}^{({\rm c})}_{I}(z)=QL^{({\rm c})}_{I}(z)Q$ 
as follows. 
\begin{eqnarray}
L^{({\rm c})}_{0}(z) A = [\tilde{H}^{({\rm c})}(z),A] \ ,
\label{lc0}
\end{eqnarray}
\begin{eqnarray}
L^{({\rm c})}_{\rm I}(z) A = [H^{({\rm c})}_{\rm I}(z),A] \ .
\label{lci}
\end{eqnarray}
Then we have
$L^{({\rm c})}(z) = L^{({\rm c})}_{0}(z) + L^{({\rm c})}_{\rm I}(z)$,
and 
the cluster memory function $\overline{G}^{({\rm c})}_{ij\sigma}(z)$ is
expressed as
\begin{eqnarray}
\overline{G}^{({\rm c})}_{ij\sigma}(z) =  
\big( A^{\dagger}_{i\sigma} \  {\big |} 
(z - \overline{L}^{({\rm c})}_{0}(z) - 
\overline{L}^{({\rm c})}_{\rm I}(z))^{-1} \,
A^{\dagger}_{j\sigma} \big) \ .
\label{gcij2}
\end{eqnarray}
Note that $\overline{L}^{({\rm c})}_{\rm I}(z)$ in the above
expression has been redefined
by Eq. (\ref{lci}), {\it i.e.}, Eq. (\ref{lic}) in which 
$\tilde{\Sigma}_{ij\sigma}(z)$ has been replaced by 
$\lambda_{ij\sigma}\tilde{\Sigma}_{ij\sigma}(z)$.

The interaction Liouvillean $\overline{L}^{({\rm c})}_{\rm I}(z)$ expands
the operator space to
\begin{eqnarray}
\overline{L}^{({\rm c})}_{\rm I}(z) |A^{\dagger}_{l \sigma}) =
\Big[ U (1- 2\langle n_{l -\sigma} \rangle ) 
- \lambda_{ll\sigma} \tilde{\Sigma}_{ll\sigma}(z) \Big] 
|A^{\dagger}_{l \sigma}) 
- \sum_{(i,j) \in {\rm c}} \Big( \delta_{lj} 
|\tilde{B}^{\dagger}_{li\sigma}) + 
\delta_{li} |\tilde{B}^{\dagger}_{lj\sigma}) 
\Big) \ , 
\label{lica}
\end{eqnarray}
\begin{eqnarray}
|\tilde{B}^{\dagger}_{lj\sigma}) &=& 
\lambda_{lj\sigma} \tilde{\Sigma}_{lj\sigma}(z) |a^{\dagger}_{j\sigma} 
\delta n_{l -\sigma})   \nonumber \\
&+&
\lambda_{lj -\sigma} \tilde{\Sigma}_{lj -\sigma}(z) |a^{\dagger}_{l\sigma} 
\delta (a^{\dagger}_{j -\sigma} a_{l -\sigma})) +
\lambda_{lj -\sigma} \tilde{\Sigma}_{lj -\sigma}(z) |a^{\dagger}_{l\sigma} 
\delta (a^{\dagger}_{l -\sigma} a_{j -\sigma})) \ .
\label{blj}
\end{eqnarray}
Therefore we introduce the projection operator
\begin{eqnarray}
\overline{P} =
\sum_{i\sigma} 
\big| A^{\dagger}_{i \sigma} \big) \chi^{-1}_{i\sigma} 
\, \big(A^{\dagger}_{i \sigma} \big|
\ ,
\label{p2} 
\end{eqnarray}
where $ \chi_{i\sigma} = \langle n_{i-\sigma} \rangle 
(1 - \langle n_{i-\sigma} \rangle)$.
It projects onto the space 
$\{ | A^{\dagger}_{i \sigma} ) \}$, while 
$\overline{Q} = 1 - \overline{P}$ eliminates the space 
$\{ | A^{\dagger}_{i \sigma} ) \}$.  Making use of these operators one
can divide the interaction Liouvillean 
$\overline{L}^{({\rm c})}_{\rm I}(z)$
into two parts.
\begin{eqnarray}
\overline{L}^{({\rm c})}_{I}(z) = \overline{P} \, 
\overline{L}^{({\rm c})}_{I}(z) 
\overline{P} + \overline{L}^{({\rm c})}_{IQ}(z) \ ,
\label{lcsepa}
\end{eqnarray}
\begin{eqnarray}
\overline{L}^{({\rm c})}_{IQ}(z) = \overline{Q} \, 
\overline{L}^{({\rm c})}_{I}(z) 
\overline{P} + \overline{L}^{({\rm c})}_{I}(z) \overline{Q} \ .
\label{lcsepa2}
\end{eqnarray}
The first term at the r.h.s. of Eq. (\ref{lcsepa}) acts in the subspace
$\{ | A^{\dagger}_{i \sigma} ) \}$, while the second term expands the
operator space beyond $\{ | A^{\dagger}_{i \sigma} ) \}$.
It should be noted that the first term at the r.h.s. of
Eq. (\ref{lcsepa2}) does not vanish in general when we take into account
the off-diagonal character of the effective medium.
This point differs from the case of the SSA~\cite{kake04-1}.

Making use of Eq. (\ref{lica}), we obtain the expression of 
$\overline{P} \, \overline{L}^{({\rm c})}_{I}(z) \overline{P}$
in Eq. (\ref{lcsepa}) as
\begin{eqnarray}
\overline{P} \, \overline{L}^{({\rm c})}_{I}(z) \overline{P} =
\sum_{ij\sigma\sigma^{\prime}} 
\big| A^{\dagger}_{i \sigma} \big) 
(\overline{\mbox{\boldmath$L$}}^{({\rm c})}_{I})_{i\sigma j\sigma^{\prime}} 
\, \big(A^{\dagger}_{j \sigma^{\prime}} \big|
\ ,
\label{plp} 
\end{eqnarray}
\begin{eqnarray}
\big( \overline{\mbox{\boldmath$L$}}^{({\rm c})}_{I}
\big)_{i\sigma j\sigma^{\prime}}
& = &
\dfrac{
\big [ U (1- 2 \langle n_{i -\sigma} \rangle) - 
\lambda_{ii\sigma} \tilde{\Sigma}_{ii\sigma}(z) \big] 
\delta_{\sigma\sigma^{\prime}} - \chi_{i\sigma}^{-1} \sum_{l(\neq i)
\in {\rm c}} 
(A^{\dagger}_{i \sigma}|\tilde{B}^{\dagger}_{il\sigma^{\prime}})
}{\chi_{i\sigma}}\delta_{ij} \hspace{10mm}  \nonumber \\
& & - 
\dfrac{(A^{\dagger}_{j \sigma}|\tilde{B}^{\dagger}_{ji\sigma^{\prime}})}
{\chi_{i\sigma}\chi_{j\sigma^{\prime}}}(1-\delta_{ij}) \ .
\label{licij}
\end{eqnarray}
Substituting Eq. (\ref{lcsepa}) into Eq. (\ref{gcij2}), we obtain
\begin{eqnarray}
\overline{G}^{({\rm c})}_{ij\sigma}(z) = 
\Big[ \overline{\mbox{\boldmath$G$}}^{({\rm c})}_{0} \cdot 
(1-\overline{\mbox{\boldmath$L$}}^{({\rm c})}_{I} \cdot 
\overline{\mbox{\boldmath$G$}}^{({\rm c})}_{0})^{-1} \Big]_{ij\sigma} \ .
\label{memcij}
\end{eqnarray}
Here the screened memory function 
$\overline{G}^{({\rm c})}_{0ij\sigma}(z) = 
(\overline{\mbox{\boldmath$G$}}^{({\rm c})}_{0})_{i\sigma j\sigma}$ 
is defined by
\begin{eqnarray}
\overline{G}^{({\rm c})}_{0ij\sigma}(z) = 
\big( A^{\dagger}_{i \sigma} {\big |}
\bigl(z - \overline{L}_{0}(z) - \overline{L}^{({\rm c})}_{IQ}(z) 
\bigr)^{-1} A^{\dagger}_{j \sigma} \big) \ .
\label{g0}
\end{eqnarray}

The expression of the cluster memory function (\ref{memcij}) is exact.
The simplest approximation is to neglect the interaction Liouville operator 
$\overline{L}^{({\rm c})}_{IQ}$ in the screened cluster memory function 
$\overline{G}^{({\rm c})}_{0ij\sigma}$, which we called the zeroth 
renormalized
perturbation theory (RPT-0).  The approximation yields in the weak 
interaction limit the correct result
of second-order perturbation theory, as well
as the exact atomic limit.  

An explicit expression of the screened cluster memory function 
in the RPT-0 can be obtained approximately as shown in Appendix.
\begin{eqnarray}
\overline{G}^{({\rm c})}_{0ij\sigma}(z) = A_{ij\sigma}
\int \dfrac{d\epsilon d\epsilon^{\prime} 
d\epsilon^{\prime\prime} 
\rho^{({\rm c})}_{ij\sigma}(\lambda, \epsilon)
\rho^{({\rm c})}_{ij-\sigma}(\lambda, \epsilon^{\prime})
\rho^{({\rm c})}_{ji-\sigma}(\lambda, \epsilon^{\prime\prime})
\chi(\epsilon, \epsilon^{\prime},
\epsilon^{\prime\prime})}
{z - \epsilon - \lambda_{\sigma}\tilde{\Sigma}_{\sigma}(\epsilon, z) 
- \epsilon^{\prime} - 
\lambda_{-\sigma}\tilde{\Sigma}_{-\sigma}(\epsilon^{\prime}, z)
+ \epsilon^{\prime\prime} + 
\lambda_{-\sigma}\tilde{\Sigma}_{-\sigma}(\epsilon^{\prime\prime}, z)}
\ ,
\label{lrpt0}
\end{eqnarray}
\begin{eqnarray}
\chi(\epsilon, \epsilon^{\prime}, 
\epsilon^{\prime\prime}) =
(1-f(\epsilon))(1-f(\epsilon^{\prime}))
f(\epsilon^{\prime\prime})
+ f(\epsilon)f(\epsilon^{\prime})
(1-f(\epsilon^{\prime\prime})) \ .
\label{chi}
\end{eqnarray}
Here $0 \le \lambda_{\sigma} \le 1$, 
and $f(\omega)$ is the Fermi distribution function.

The prefactor $A_{ij\sigma}$ in Eq. (\ref{lrpt0}) 
has been introduced to recover the exact second moment of the spectrum.
\begin{eqnarray}
A_{ij\sigma} = \dfrac{\chi_{i\sigma}}
{\langle n_{i -\sigma} \rangle_{\rm c} 
(1 - \langle n_{i -\sigma} \rangle_{\rm c})}\delta_{ij} + 1 -
\delta_{ij} 
\ .
\label{aij}
\end{eqnarray}
Here the electron number for a cavity state 
$\langle n_{i\sigma} \rangle_{\rm c}$ is defined by
\begin{eqnarray}
\langle n_{i\sigma} \rangle_{\rm c} = \int d\epsilon 
\rho^{({\rm c})}_{ii\sigma}(\lambda, \epsilon) f(\epsilon) \ .
\label{cavn}
\end{eqnarray}
The densities of states for the cavity state are given by
\begin{eqnarray}
\rho^{({\rm c})}_{ij\sigma}(\lambda, \epsilon) = 
-\frac{1}{\pi} {\rm Im} \ [(
\mbox{\boldmath$F$}_{\rm c}(\lambda, z)^{-1}
+ \overline{\lambda}_{\sigma}
\tilde{\mbox{\boldmath$\Sigma$}}^{({\rm c})}(z))^{-1}]_{ij\sigma} \ ,
\label{rhocav}
\end{eqnarray}
\begin{eqnarray}
(\mbox{\boldmath$F$}_{\rm c}(\lambda, z))_{ij\sigma} = 
F_{ij\sigma}(z) = 
\sum_{k} \dfrac{\langle i|k \rangle \langle k|j \rangle}
{z - \epsilon_{k\sigma} - 
\overline{\lambda}_{\sigma} \tilde{\Sigma}_{k\sigma}(z)} \ .
\label{fcl}
\end{eqnarray}
Furthermore, $\overline{\lambda}_{\sigma} = 1 - \lambda_{\sigma}$, 
$(\tilde{\mbox{\boldmath$\Sigma$}}^{({\rm c})}(z))_{ij\sigma} = 
\tilde{\Sigma}_{ij\sigma}(z) \ \ (i,j \in {\rm c})$, and 
$\tilde{\Sigma}_{k\sigma}(z)$ is the Fourier transform of 
$\tilde{\Sigma}_{ij\sigma}(z)$;
$\tilde{\Sigma}_{k\sigma}(z) = \sum_{j}
\tilde{\Sigma}_{j0\sigma}(z) \exp (i\mbox{\boldmath$k$} \cdot
\mbox{\boldmath$R$}_{j})$.

The simplified self-energy $\tilde{\Sigma}_{\sigma}(\epsilon, z)$ 
in Eq. (\ref{lrpt0}) is calculated from 
$\tilde{\Sigma}_{ij\sigma}(z)$ as
\begin{eqnarray}
\tilde{\Sigma}_{\sigma}(\epsilon, z) =
\sum_{i} \dfrac{\rho^{0}_{ij}(\epsilon-\epsilon_{\sigma})}
{\rho^{0}(\epsilon-\epsilon_{\sigma})} \tilde{\Sigma}_{ij\sigma}(z) \ ,
\label{peffk2}
\end{eqnarray}
where ${\rho^{0}_{ij}(\epsilon)}$ is the DOS of the noninteracting system
defined by ${\rho^{0}_{ij}(\epsilon)} = \sum_{k} \langle i|k \rangle 
\delta(\epsilon - \epsilon_{k}) \langle k|j \rangle$
and ${\rho^{0}(\epsilon)}$ is defined by  
$\rho^{0}(\epsilon) = 1/N \ \sum_{k}  
\delta(\epsilon - \epsilon_{k})$.

Finally the simplified cluster memory function in the RPT-0 is given by
Eq. (\ref{memcij}), Eq. (\ref{licij}) in which 
$\{ \lambda_{ij\sigma} \}$ has been replaced by 
$\{ \lambda_{\sigma} \}$, and Eq. (\ref{lrpt0}).
In the present scheme, we first assume $\tilde{\Sigma}_{ij\sigma}(z)$.
Then we calculate the coherent Green function $F_{ij\sigma}(z)$
(Eq. (\ref{fcl})), the screened cluster memory function 
$\overline{G}^{(c)}_{0ij\sigma}(z)$ according to Eq. (\ref{lrpt0}), 
as well as the atomic
frequency matrix $\big( \overline{\mbox{\boldmath$L$}}^{({\rm c})}_{I}
\big)_{i\sigma j\sigma^{\prime}}$ given by Eq. (\ref{licij}). 
Using these matrices, we calculate the 
cluster memory function $\overline{G}^{(c)}_{ij\sigma}(z)$ 
(Eq. (\ref{memcij})).
Note that the static quantities 
$(A^{\dagger}_{i\sigma^{\prime}}|
\tilde{B}^{\dagger}_{jl\sigma^{\prime}})$ in Eq. (\ref{licij}) 
have to be calculated separately, for example, by means of the local
ansatz wavefunction method~\cite{fulde}.  
After having calculated cluster memory functions,
we can obtain the memory functions (\ref{icrii}) and (\ref{icrij})
according to the incremental scheme.  Then we calculate the diagonal and
off-diagonal self-energies (\ref{self}).
If the self-consistent condition (\ref{sceqij}) is not satisfied, we
repeat the above-mentioned procedure renewing the medium 
$\tilde{\Sigma}_{ij\sigma}(z)$ until the self-consistency (\ref{sceqij})
is achieved.
When we obtain the self-consistent solution 
$\tilde{\Sigma}_{ij\sigma}(z)$, we can calculate the
excitation spectrum $\rho_{k\sigma}(\omega)$ from
the Green function (\ref{gk}) and the DOS $\rho_{\sigma}(\omega)$ from
Eq. (\ref{rhoii}). 
\vspace{2mm}

\section{Nonlocal Excitations on a Simple-Cubic Lattice}
We present here the numerical results of excitation spectra of the
Hubbard model on a simple cubic lattice at half-filling in the
paramagnetic state in order to examine the nonlocal correlations in the
FSCPM.   
As in our previous calculations,
we choose the parameters $\lambda_{\sigma}=0$ in Eq. (\ref{lrpt0}); 
we start from the
cavity cluster state ($\lambda_{ij}=0$) for the calculation of the
memory function.  The form (\ref{lrpt0}) with $\lambda_{\sigma}=0$
reduces to the iterative pertubation scheme~\cite{kaju96} 
at half-filling in infinite dimensions. 
Note that we need not to calculate
$\tilde{\Sigma}_{\sigma}(\epsilon, z)$ defined by Eq. (\ref{peffk2}) as
well as $(A^{\dagger}_{j\sigma}|\tilde{B}^{\dagger}_{il\sigma})$ in
Eq. (\ref{licij}) when $\lambda_{\sigma}=0$.

We adopt the nearest-neighbor transfer integral $t$
here and in the following, choose the energy unit as $|t|=1$.  The Fourier
transform of the transfer integrals is given by 
$\epsilon=-2(\cos k_{x} + \cos k_{y} + \cos k_{z})$ in the unit of
lattice constant $a$.
Furthermore, in Eqs. (\ref{icrii}) and (\ref{icrij}) 
we take into account single-site and pair-site terms
only, but the latters up to the 10th nearest neighbors.
In the FSCPM, we have to perform a numerical $k$ integration to obtain
the coherent Green function $F_{ij\sigma}$ (see Eq. (\ref{fcl})).
We have adopted a $80 \times 80 \times 80$ mesh in the first Brillouin
zone for such integration. 

In the numerical calculations of the screened memory function
(\ref{lrpt0}), we applied the method of Laplace 
transformation~\cite{schweitz91}.  This
reduces the 3-fold integrals with respect to energy into the one-fold
integral with respect to time.
%
\begin{figure}[tb]
\includegraphics{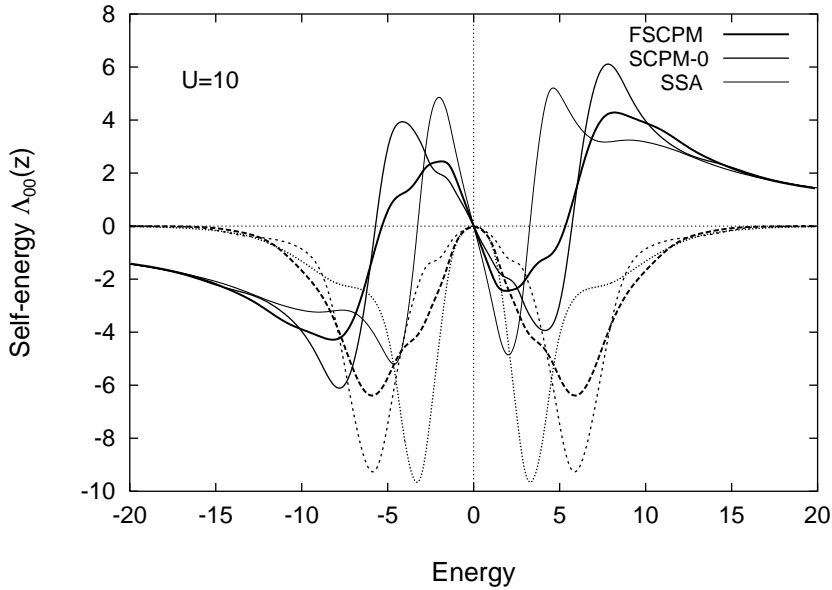}%
\caption{\label{figself00}
Diagonal self-energies for $U=10$ in the FSCPM (thick curves), the SCPM-0
 (middle-size curves), and the SSA (thin curves).
The real (imaginary) part in each method is drawn by the solid
 (dotted) curve.
}
\end{figure}
%
%
\begin{figure}[tb]
\includegraphics{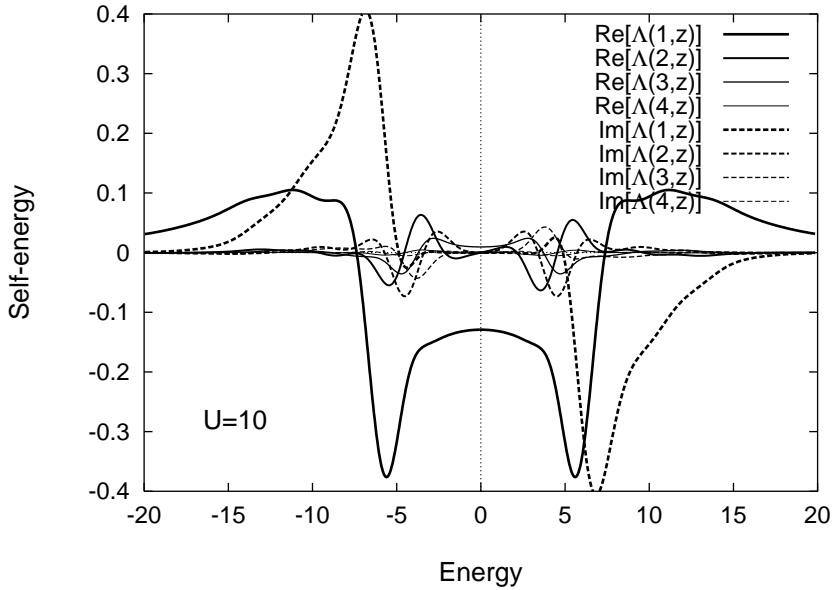}%
\caption{\label{figselfj0}
Off-diagonal self-energy $\Lambda(n,z)$ for the $n$-th nearest neighbor
 pairs ($n=1,2,3,4$).
}
\end{figure}
%

Starting from an initial set of values $\{ \tilde{\Sigma}_{0j}(z)=0 \}
(j=0, 1, \cdots, M)$, we calculated  
$\{ F_{ij}(z) \}$ (Eq. (\ref{fcl})), 
$\{ \rho^{(c)}_{ij}(z) \}$ (Eq. (\ref{rhocav})),
$\{ \overline{G}^{(\rm c)}_{0ij}(z) \}$ (Eq. (\ref{lrpt0})),
$\{ \overline{G}^{(\rm c)}_{ij}(z) \}$ (Eq. (\ref{memcij})), and
$\{ \overline{G}_{ij}(z) \}$ (Eqs. (\ref{icrii}) and (\ref{icrij})), 
and finally obtained
$\Lambda_{ij\sigma}(z) = U^{2} \overline{G}_{ij}(z)$.
We have repeated the same procedure until self-consistency (\ref{sceqij})
was achieved.

We show in Fig. \ref{figself00} the self-consistent diagonal self-energy 
at $U=10$ and compare the results with the SSA and SCPM-0 calculations.
The SCPM-0 shifts the spectral weight of the SSA towards the higher energy
region.  The FSCPM suppresses the amplitude of the SCPM-0 self-energy in the
high energy region.  In the low energy region, we find that the
imaginary part is close to that of the SSA, while the real part is
in-between the SSA and the SCPM-0.

The off-diagonal self-energies are shown in Fig. \ref{figselfj0}.
We find again that self-energies are suppressed by the full self-consistency.
Furthermore the self-energies rapidly damp with increasing interatomic
distance.  The fourth-nearest neighbor contribution and the
contribution from more distant pairs can be neglected when $U=10$, though
we took into account the off-diagonal ones up to 10th nearest neighbors. 

\begin{figure}[tb]
\includegraphics{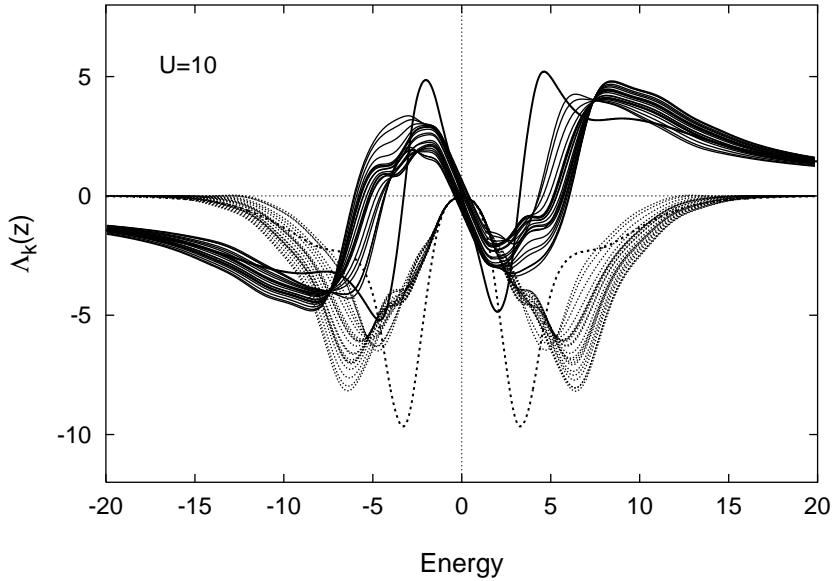}%
\caption{\label{figskk}
Momentum dependent self-energies at various $k$ points along the high
 symmetry line $\Gamma$-X-M-R.  The real (imaginary) parts are 
shown by solid (dotted) curves.  Note that the real part at the $\Gamma$ 
point shows a maximum at energy $\omega \approx 8$ and a second maximum
 at $\omega \approx -3$, while the imaginary part shows a minimum at
 $\omega \approx 6$ and a second minimum at $\omega \approx -5$.  The 
results of the SSA are also shown by the thick solid and dotted curves.
}
\end{figure}

The momentum-dependent self-energies $\Lambda_{k\sigma}(z)$ are also
suppressed by the full self-consistency. 
As shown in Fig. \ref{figskk}, in the low energy region the imaginary 
parts of $\Lambda_{k\sigma}(z)$ 
hardly depend on momentum $\mbox{\boldmath$k$}$ 
and are close to those of the SSA.
In the incoherent region $|\omega| \gtrsim 2$, 
both real and imaginary parts show considerable
$\mbox{\boldmath$k$}$ dependence.  For example, the imaginary part of
$\Lambda_{k\sigma}(z)$ shows at the  $\Gamma$ point a minimum at 
$\omega \approx 6$
and a second one at $\omega \approx -5$, while
it shows a minimum at $\omega = -6$ and the second one at 
$\omega = 5$ at the R point.
%
%

\begin{figure}[tb]
\includegraphics{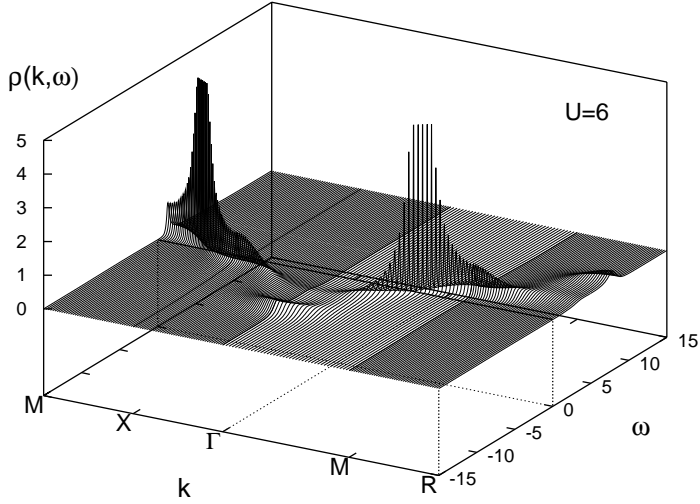}%
\caption{\label{figdosk6}
Single-particle excitation spectra along the high-symmetry lines at
 $U=6$.  The Fermi level is indicated by a bold line.
}
\end{figure}

\begin{figure}[tb]
\includegraphics{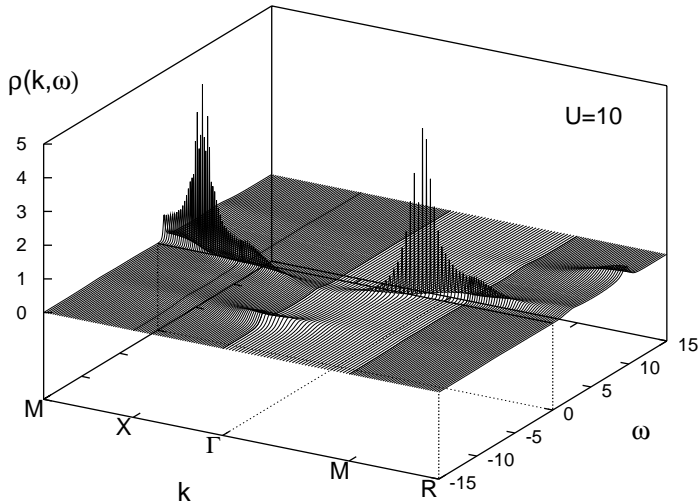}%
\caption{\label{figdosk10}
Single-particle excitation spectra at $U=10$.
}
\end{figure}

\begin{figure}[tb]
\includegraphics{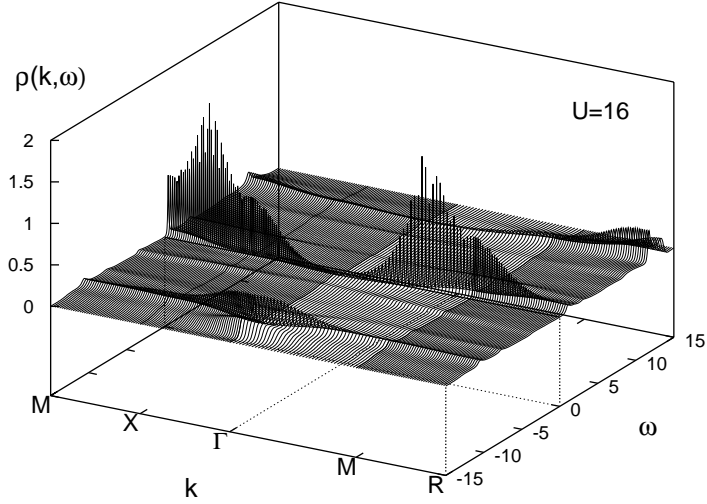}%
\caption{\label{figdosk16}
Single-particle excitation spectra at $U=16$. Note that the vertical
 scale has been changed in order to show detailed structure of 
excitations.
}
\end{figure}

Calculated momentum dependent excitation spectra are shown in Figs. 
\ref{figdosk6}, \ref{figdosk10}, and \ref{figdosk16}. 
For rather weak Coulomb interaction ($U=6$), we find that 
the quasiparticle band is
reduced by about 30\% in width as compared with that of the
noninteracting band.  As seen in Fig. \ref{figdosk6},
the Mott-Hubbard incoherent bands appear at
$|\omega| \approx 7.5$ around the $\Gamma$ and R point. 
The spectral weight of the Mott-Hubbard bands 
is reduced by 30\% when it is compared with that of the SCPM-0.
%
%
%
%

In the intermediate Coulomb interaction regime ($U=10$), the
quasiparticle band width becomes narrower and is reduced by 25\% as
compared with that of the SCPM-0.  The spectral weight of the
quasiparticle band around the $\Gamma$ and R point moves farther 
to the lower and upper Mott-Hubbard bands.  The latters are more 
localized in the vicinity of the $\Gamma$ and R point than in the
SCPM-0, and their peaks are reduced by 25\% as compared with those in the
SCPM-0.
%
%
%
\begin{figure}[tb]
\includegraphics{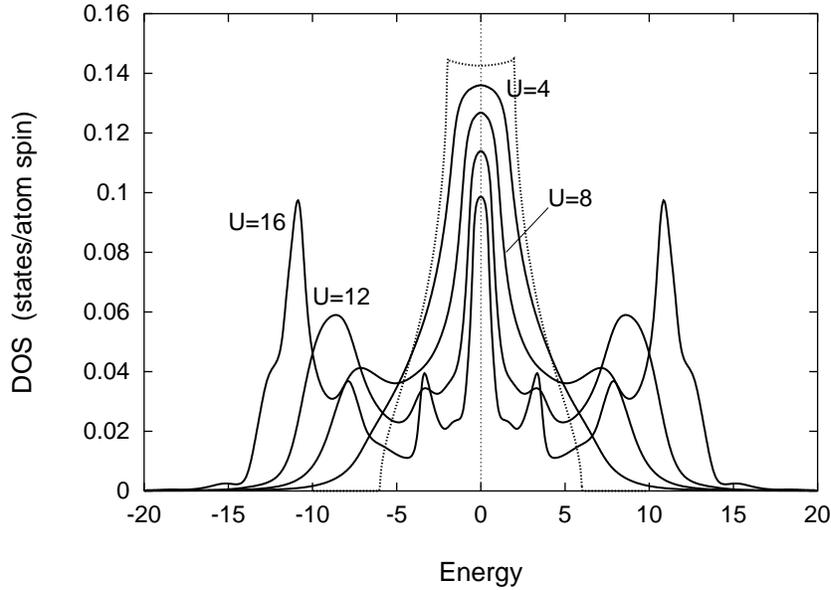}%
\caption{\label{figdosu}
The total DOS for $U=4, 8, 12, 16$.  The DOS for noninteracting system
 is shown by dotted curve.
}
\end{figure}
%

In the FSCPM scheme, we can obtain the momentum-dependent spectra in the
strong Coulomb interaction regime ($U=16$).  This is shown 
in Fig. \ref{figdosk16}.  In this regime, the
quasiparticle band becomes narrower and the spectral weight becomes
smaller.  New excitations which appear in this regime are the
subbands at $|\omega| \approx 3.3$ showing weak momentum dependence.  
Because of the strong AF intersite
correlations at half-filling, these excitations may be interpreted as
``shadow'' bands due to AF correlations, whose excitation spectra
in a simple spin density wave (SDW) picture\cite{grober00}
may be given by
$\epsilon^{SDW}_{\pm}(k) = \pm \sqrt{\tilde{\epsilon}^{2}_{k}+\Delta^{2}}$. 
Here
$\tilde{\epsilon}_{k}$ is a quasiparticle band, $\Delta$ is an exchange
splitting defined by $\Delta = U_{\rm eff} |m|/2$, $|m|$ being a
temporal amplitude of the magnetic moments with long-range AF
correlations, and $U_{\rm eff}$ is an effective Coulomb
interaction.

As seen in Fig. \ref{figdosk16}  
the Mott-Hubbard bands in the strong Coulomb interaction regime 
show a weak momentum dependence around $|\omega| \approx \pm 11$.
It should be noted that the splitting between the upper and lower
Hubbard bands is about
22, which is larger than $U=16$ (a value yielding the atomic limit).
In the strong Coulomb interaction limit at half-filling, we expect
strong AF intersite correlations due to the super-exchange interaction
$J=4|t|^{2}/U$.  The energy to remove (add) electrons is then expected
to be $\epsilon_{0} - z_{\rm NN}|J|$ ($\epsilon_{0} + U + z_{\rm NN}|J|$), 
$z_{\rm NN}$ being the number of nearest neighbors ($z_{\rm NN}=6$ 
in the present case).
Thus the splitting is expected to be $U + 2z_{\rm NN}|J|$ instead of $U$.  
This formula yields the splitting 19 instead of $U=16$.
The former seems to be consistent with the value 22.
We also find additional excitations at $|\omega| \approx 8$.  These
sub-bands may be interpreted as local excitations of the lower and upper
Hubbard bands without AF correlations.
%
\begin{figure}[tb]
\includegraphics{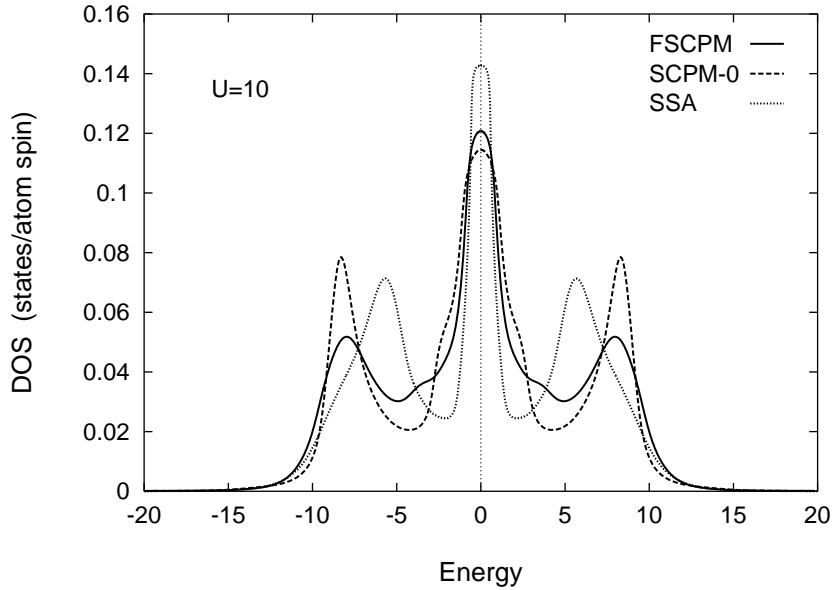}%
\caption{\label{figdosu10}
The DOS in the FSCPM (solid curve), the SCPM-0 (dashed curve), and the
 SSA (dotted curve) at $U=10$.
}
\end{figure}
%

The total densities of states are presented in Fig. \ref{figdosu} 
for various values of $U$.  
For weak Coulomb interactions $U \lesssim
6$, the deviation of the spectra from those of the SCPM-0 is negligible.
When $U > 6$, Mott-Hubbard bands appear and the DOS deviate from the
SCPM-0.  An example is shown in Fig. \ref{figdosu10} for $U=10$.  
We find there that the full self-consistency
suppresses the weight of the Mott-Hubbard bands and enhances the
quasiparticle peaks.  Resulting DOS is in-between the SSA and the SCPM-0.
When $U > 10$, a shadow band develops around $|\omega| = 3$ as seen
in Fig. \ref{figdosu}.  Furthermore, for $U > 12$ we find two peaks 
in each Mott-Hubbard band; one is due to the excitations 
without intersite AF correlations, another is due to the excitations 
with strong AF correlations.

\begin{figure}[tb]
\includegraphics{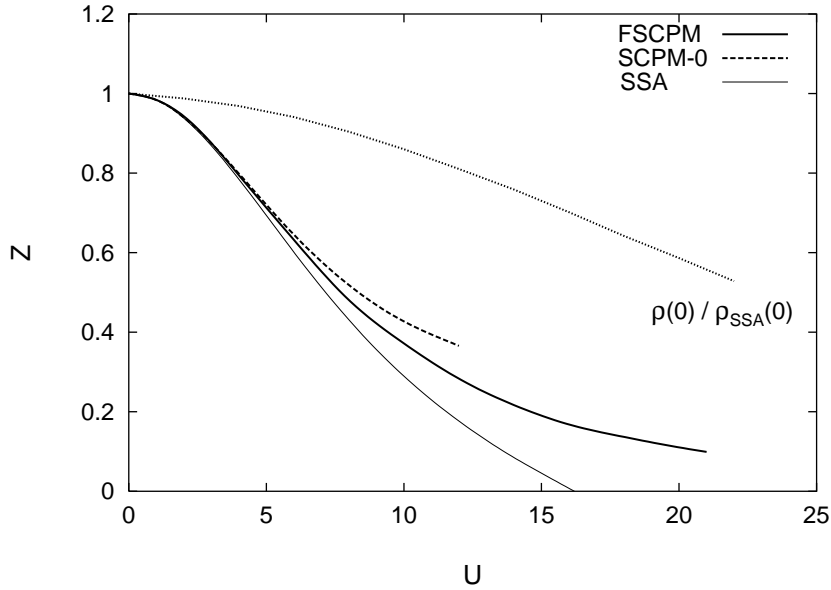}%
\caption{\label{figzu}
The average quasiparticle weight vs. Coulomb interaction curves in the
 FSCPM (solid curves), the SCPM-0 (dashed curve), and the SSA (thin solid
 curve). The DOS divided by the SSA one at the Fermi level is shown
 by dotted curve.  Note that $\rho_{\rm SSA}(0)$ equals the
 noninteracting DOS $\rho^{0}(0)$.  $\rho(0)/\rho_{\rm SSA}(0)$ beyond
 $U_{\rm c}(\rm SSA)=16.0$ stands for $\rho(0)/\rho^{0}(0)$.
}
\end{figure}

A remarkable point of the nonlocal excitation spectra is that the
quasiparticle peak at the Fermi level reduces with increasing Coulomb
interaction $U$.  As shown in Fig \ref{figzu}, the ratio 
of $\rho(0)/\rho_{\rm SSA}(0)$ to the SSA, monotonically 
decreases with increasing $U$, and reaches 0.59 when $U=20$.
%
%
%
\begin{figure}[tb]
\includegraphics{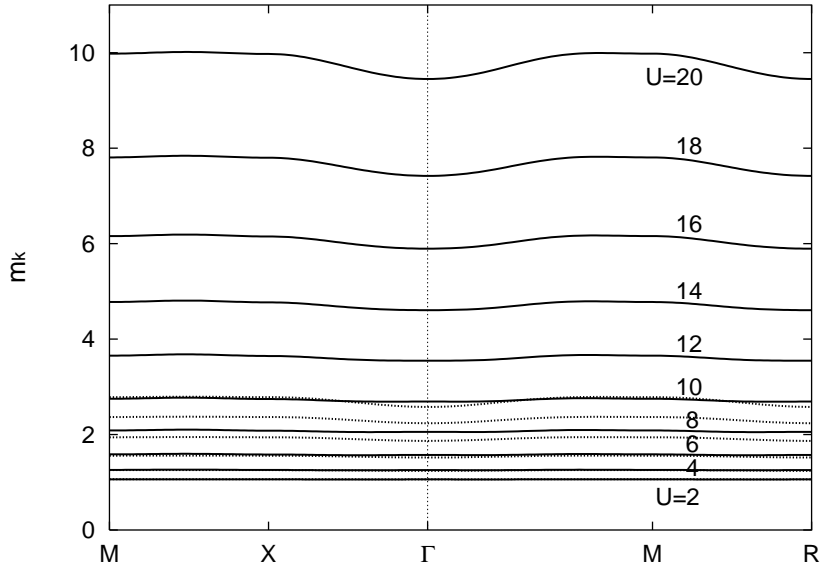}%
\caption{\label{figmk}
The momentum-dependent effective mass curves along the high-symmetry
 lines for various Coulomb interaction $U$.  
The curves in the SCPM-0 are also shown by the
 dotted curves for $U=2,4,6,8,10$, and $12$ from the bottom to the top.
}
\end{figure}
%

Momentum-dependent effective masses $m_{k}$ along the high symmetry line
are presented in Fig. \ref{figmk}.  In the FSCPM calculations, we 
obtained the self-consistent
$m_{k}$ up to $U=20$.  We find that $m_{k}$ are
approximately equal to those in the SCPM-0 for $U \lesssim 6$.  But, for 
$U > 6$, the FSCPM enhances the mass $m_{k}$ as compared with those of
the SCPM-0.
Momentum dependence of $m_{k}$ becomes larger with increasing $U$.  The
mass has a minimum ({\it e.g.}, 9.4 for $U=20$) at 
$\Gamma (0, 0, 0)$ and R $(\pi, \pi, \pi)$ points, and has a maximum 
({\it e.g.}, 10.0 for $U=20$) at $(\pi, \pi/2, 0)$.

The average quasiparticle weight $Z=(N^{-1} \sum_{k} m_{k})^{-1}$ vs.
$U$ curve is shown in Fig. \ref{figzu}.
Quasiparticle weight in the SSA monotonically decreases and vanishes for
$U_{c}=16.0$.  The curve in the SCPM-0 deviates upwards from the SSA.
The curve in the FSCPM deviates downwards 
from the SCPM-0 beyond $U \approx 5$,
and is between the SSA and the SCPM-0.  The present result suggests
that the critical Coulomb interaction for the divergence of the
effective mass is $U_{\rm c}(m^{\ast}=\infty) \gtrsim 30$.
We want to mention that for the Gutzwiller wave 
function a critical Coulomb interaction 
$U_{\rm c}(m^{\ast}=\infty)$
exists only in infinite dimensions 
({\it i.e.}, in the SSA)~\cite{gutz63,yoko87,bri70}.

\begin{figure}[tb]
\includegraphics{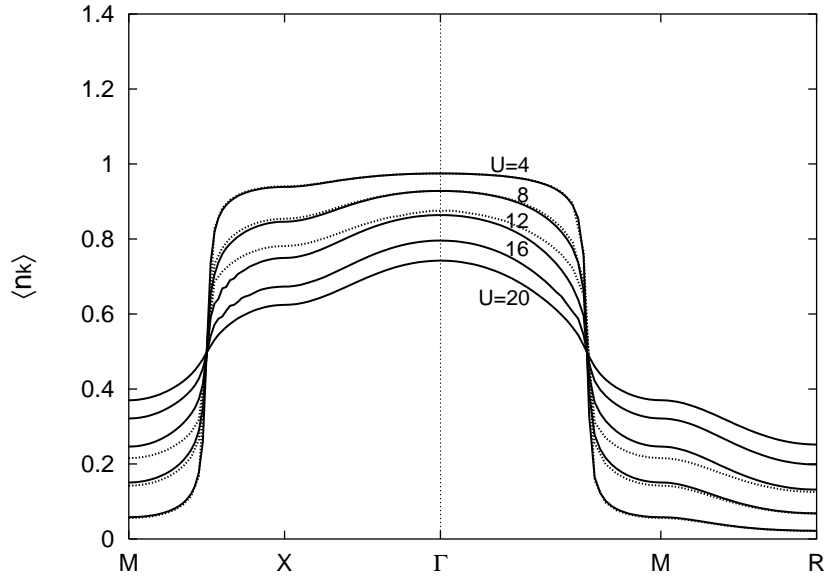}%
\caption{\label{fignk}
Momentum distribution along the high-symmetry lines in the FSCPM (solid
 curves).  The curves in the SCPM-0 are also drawn by dotted curves for
 $U=4,8,12$.
}
\end{figure}

We have also calculated the momentum distribution as shown in
Fig. \ref{fignk}.   The distributions show a jump at the Fermi surface, and
extend beyond the surface.  The basic behavior of the distribution is
similar to that in the SCPM-0.  For $U \lesssim 6$, the curves $\langle
n_{k} \rangle$ agree with those of the SCPM-0.  
When $U > 6$, the FSCPM reduces $\langle n_{k} \rangle$
of the SCPM-0 below the Fermi level, and enhances it 
above the Fermi level suggesting increased localization due to full 
self-consistency.  
%
%

\section{Summary and Discussions}
We have developed a fully self-consistent projection operator method
(FSCPM) for nonlocal excitations by using the projection
technique for the retarded Green function and the effective medium.  
The method makes use of an energy-dependent Liouville operator
$\tilde{L}(z)$ with a corresponding Hamiltonian for an off-diagonal
effective medium $\tilde{\Sigma}_{ij\sigma}(z)$.  It allows us to 
calculate the nonlocal self-energy $\Lambda_{ij\sigma}(z)$ by using the
incremental cluster expansion from the off-diagonal medium.  
Each term of the expansion 
is calculated from the memory function for clusters
in a nonlocal effective medium.  
The latters are obtained by the renormalized perturbation theory within
the RPT-0.
The off-diagonal effective medium
$\tilde{\Sigma}_{ij\sigma}(z)$ is determined from a fully self-consistent
condition $\{ \tilde{\Sigma}_{ij\sigma}(z)=\Lambda_{ij\sigma}(z) \}$.
It is a generalization of the CPA as given in Eq. (\ref{gcpa}), {\it i.e.},
$\big( A^{\dagger}_{i\sigma} \  {\big |} 
\overline{G}_{0} T \overline{G}_{0} A^{\dagger}_{j\sigma} \big) = 0$
where $T$ denotes a scattering $T$-matrix from the off-diagonal medium and 
$\overline{G}_{0}$ is a resolvent for the Liouville operator 
$\overline{L}_{0}$ corresponding to the medium.
In this way we can take into account the long-range intersite
correlations as extensive as we want in each order of expansion.  We obtain
momentum-dependent spectra of high resolution from the weak
to the strong Coulomb interaction regime.

The present theory reduces to the PM-CPA when we omit all off-diagonal
matrix elements of the effective medium 
$\tilde{\Sigma}_{ij\sigma}(z) (i \neq j)$ and 
all the off-diagonal self-energy contributions.
The theory reduces to
a self-consistent theory (SCPM-0) when we omit the off-diagonal medium
$\tilde{\Sigma}_{ij\sigma}(z) (i \neq j)$, but take into account all
the off-diagonal self-energy contribution
$\Lambda_{ij\sigma}(z)$.

We have performed numerical calculations of the excitation spectra for
the half-filled Hubbard model on a simple cubic lattice by using the 
FSCPM within the two-site approximation.
We have obtained the self-consistent nonlocal self-energy up to the Coulomb 
interaction $U \approx 20$.  We found that the FSCPM suppresses
the amplitudes of local and nonlocal self-energy
$\Lambda_{ij\sigma}(z)$ for an intermediate strength of Coulomb
interactions, reduces the weight of the Mott-Hubbard bands, and enhances
the quasiparticle peaks at the Fermi level in the average DOS when
they are compared with those in the SCPM-0.  
Moreover the FSCPM enhances the momentum-dependent effective mass $m_{k}$ as
compared with the SCPM-0.  Thus the $Z-U$ curve is located between
the SSA and the SCPM-0.  
These results indicate that the full self-consistency tends to suppress
the nonlocal effects found in the SCPM-0.
We suggest a critical Coulomb interaction
$U_{\rm c}(m^{\ast}=\infty) \gtrsim 30$ for the simple cubic lattice,
which is much larger than the SSA value $U_{\rm c}=16$.
In order to obtain the explicit value of $U_{\rm c}$, we have to take
into account larger clusters embedded in the off-diagonal medium.

The FSCPM enables us to investigate the nonlocal excitation spectra in the
strongly correlated region.  There the quasiparticle bands becomes
narrower and their weight becomes smaller.  
We found shadow band excitations at $|\omega| \approx 3$ 
due to strong AF correlations
and Mott-Hubbard sub-bands at $|\omega| \approx U/2$ 
without intersite correlations.
Moreover, we found that strong AF correlations
can shift the Mott-Hubbard bands towards higher energies.

In the present calculations, we investigated the effects of nonlocal
correlations assuming the paramagnetic state. One has to extend the
calculations to the AF case in the next step because the ground state
of the three dimensional Hubbard model is believed to be antiferromagnetic
in general.  Furthermore, the numerical results of calculations 
in the strongly correlated region
presented here should be extended by taking into
account the contributions from larger clusters in the incremental
cluster expansion.  Improvements of the self-energies for the
clusters in the off-diagonal effective medium are also left for future
investigations towards more quantitative calculations of the 
nonlocal excitations.

\section*{Acknowledgment}
This work was supported by Grant-in-Aid for Scientific Research (19540408). 
Numerical calculations have been partly carried out with use of 
the Hitachi SR11000 in the Supercomputer Center, Institute of 
Solid State Physics, University of Tokyo.
\vspace{10mm}
\clearpage

\appendix
\section{Derivation of Approximate Expression for the Screened Cluster
Memory Function (\ref{lrpt0})}   
\vspace{5mm}

An approximate expression (\ref{lrpt0}) of the screened
cluster memory function in the RPT-0 can be obtained as follows 
by using the momentum representation of the operator space.  

For that purpose, we first express the local
operator $A^{\dagger}_{i\sigma}$ by means of the creation and
annihilation operators in the momentum representation as
\begin{eqnarray}
A^{\dagger}_{i \sigma} = \sum_{k,k^{\prime},k^{\prime\prime}}
a^{\dagger}_{k \sigma} \delta 
(a^{\dagger}_{k^{\prime} -\sigma}a_{k^{\prime\prime} -\sigma})
\langle k|i \rangle \langle k^{\prime}|i \rangle 
\langle i|k^{\prime\prime} \rangle \ .
\label{aif}
\end{eqnarray}
Here
$\langle k|i \rangle = \langle i|k \rangle^{\ast} = 1/\sqrt{N} 
\ \exp (i\mbox{\boldmath$k$}\cdot\mbox{\boldmath$R$}_{i})$.
The operator $a^{\dagger}_{k \sigma} \delta 
(a^{\dagger}_{k^{\prime} -\sigma}a_{k^{\prime\prime} -\sigma})$ in 
Eq. (\ref{aif}) is the eigen state of 
$\overline{L}_{0}(z)=Q\tilde{L}(z)Q$, {\it i.e.}, 
\begin{eqnarray}
\overline{L}_{0}(z)|a^{\dagger}_{k \sigma} \delta 
(a^{\dagger}_{k^{\prime} -\sigma}a_{k^{\prime\prime} -\sigma})) &=&
[\epsilon_{\sigma} + \epsilon_{k} + \tilde{\Sigma}_{k\sigma}(z) 
+ \epsilon_{k^{\prime}} + \tilde{\Sigma}_{k^{\prime} -\sigma}(z)
 \hspace{20mm}  \nonumber \\
& & \hspace*{20mm} - \epsilon_{k^{\prime\prime}} - 
\tilde{\Sigma}_{k^{\prime\prime} -\sigma}(z)]
|a^{\dagger}_{k \sigma} \delta 
(a^{\dagger}_{k^{\prime} -\sigma}a_{k^{\prime\prime} -\sigma})) \ . 
\label{l0eigen}
\end{eqnarray}
Here 
$\tilde{\Sigma}_{k\sigma}(z)$ is the Fourier transform of
$\tilde{\Sigma}_{ij\sigma}(z)$, which is defined by
$\tilde{\Sigma}_{k\sigma}(z) = \sum_{j}
\tilde{\Sigma}_{j0\sigma}(z) \exp (i\mbox{\boldmath$k$} \cdot
\mbox{\boldmath$R$}_{j})$.

Substituting Eq. (\ref{aif}) into the approximate expression 
$\overline{G}^{({\rm c})}_{0ij\sigma}(z) = 
\big( A^{\dagger}_{i \sigma} \big|
\big( z - \overline{L}_{0}(z) \big)^{-1} A^{\dagger}_{j \sigma} \big)$
and making use of the relation (\ref{l0eigen}), we reach the explicit 
expression of the screened cluster memory function in the RPT-0 as
\begin{eqnarray}
\overline{G}^{({\rm c})}_{0ij\sigma}(z) =
\sum_{k k^{\prime} k^{\prime\prime} k_{1} k_{1}^{\prime}
k_{1}^{\prime\prime}}
\langle i|k_{1} \rangle \langle i|k^{\prime}_{1} \rangle
\langle k^{\prime\prime}_{1}|i \rangle \langle k^{\prime\prime}_{1}|i \rangle
(\overline{\mbox{\boldmath$G$}}^{({\rm c})}_{0}
)_{k_{1}k_{1}^{\prime}k_{1}^{\prime\prime}\sigma 
k k^{\prime}k^{\prime\prime}\sigma}
\langle k|j \rangle \langle k^{\prime}|j \rangle 
\langle j|k^{\prime\prime} \rangle \ ,
\label{g0ij2}
\end{eqnarray}
\begin{eqnarray}
\overline{\mbox{\boldmath$G$}}^{({\rm c})}_{0} = \mbox{\boldmath$\chi$}
( z - \overline{\mbox{\boldmath$L$}}_{0} - \mbox{\boldmath$v$}_{\rm c}
)^{-1} \ ,
\label{g0matrix}
\end{eqnarray}
\begin{eqnarray}
(\mbox{\boldmath$\chi$})_{k_{1} k^{\prime}_{1} k^{\prime\prime}_{1}
\sigma^{\prime} k k^{\prime} k^{\prime\prime}\sigma} =
(a^{\dagger}_{k_{1} \sigma^{\prime}} \delta 
(a^{\dagger}_{k^{\prime}_{1} -\sigma^{\prime}}
a_{k^{\prime\prime}_{1} -\sigma^{\prime}})
|a^{\dagger}_{k \sigma} \delta 
(a^{\dagger}_{k^{\prime} -\sigma}a_{k^{\prime\prime} -\sigma})) \ ,
\label{chi0matrix}
\end{eqnarray}
\begin{eqnarray}
(\overline{\mbox{\boldmath$L$}}_{0})_{k_{1} k^{\prime}_{1} k^{\prime\prime}_{1}
\sigma^{\prime} k k^{\prime} k^{\prime\prime}\sigma} &=&
\big( \epsilon_{\sigma} + \epsilon_{k} + \tilde{\Sigma}_{k\sigma}(z) 
+ \epsilon_{k^{\prime}} + \tilde{\Sigma}_{k^{\prime} -\sigma}(z)
 \hspace{10mm}  \nonumber \\
& & \hspace*{30mm} - \epsilon_{k^{\prime\prime}} - 
\tilde{\Sigma}_{k^{\prime\prime} -\sigma}(z) \big) 
\delta_{k_{1}k}\delta_{k^{\prime}_{1}k^{\prime}}
\delta_{k^{\prime\prime}_{1}k^{\prime\prime}} 
\delta_{\sigma\sigma^{\prime}} \ ,
\label{l0matrix}
\end{eqnarray}
\begin{eqnarray}
(\mbox{\boldmath$v$}_{\rm c})_{k_{1} k^{\prime}_{1} k^{\prime\prime}_{1}
\sigma^{\prime} k k^{\prime} k^{\prime\prime}\sigma} &=&
\sum_{lm \in {\rm c}} \Big[
- \overline{\lambda}_{lm\sigma}\tilde{\Sigma}_{lm\sigma}(z)
\langle k_{1}|l \rangle \langle m|k \rangle
\delta_{k_{1}^{\prime} k^{\prime}}\delta_{k^{\prime\prime}_{1} 
k^{\prime\prime}}  \nonumber \\
& & \hspace*{10mm} 
- \overline{\lambda}_{lm -\sigma}\tilde{\Sigma}_{lm -\sigma}(z)
\langle k_{1}^{\prime}|l \rangle \langle m|k^{\prime} \rangle
\delta_{k_{1} k}\delta_{k^{\prime\prime}_{1} k^{\prime\prime}}
\nonumber \\
& & \hspace*{10mm}
+ \overline{\lambda}_{lm -\sigma}\tilde{\Sigma}_{lm-\sigma}(z)
\langle k^{\prime\prime}|l \rangle \langle m|k^{\prime\prime}_{1} \rangle
\delta_{k_{1} k}\delta_{k^{\prime}_{1} k^{\prime}}
\Big] \delta_{\sigma\sigma^{\prime}} \ .
\label{vcmatrix}
\end{eqnarray}

The screened memory function (\ref{g0ij2}) depends on the choice of 
$\{ \lambda_{ij\sigma} \}$.  When we choose $\{ \lambda_{ij\sigma}=1 \}$
and adopt the Hartree-Fock approximation when computing the static 
average in Eq. (\ref{chi0matrix}), we have 
\begin{eqnarray}
\overline{G}^{({\rm c})}_{0ij\sigma}(z) =
\sum_{k, k^{\prime}, k^{\prime\prime}}
\dfrac{ \langle i|k \rangle \langle i|k^{\prime} \rangle
\langle k^{\prime\prime}|i \rangle 
\chi(\epsilon_{k\sigma}, \epsilon_{k^{\prime}-\sigma}, 
\epsilon_{k^{\prime\prime}-\sigma})
\langle k|j \rangle \langle k^{\prime}|j \rangle 
\langle j|k^{\prime\prime} \rangle }
{ z -
\epsilon_{k\sigma} - \tilde{\Sigma}_{k\sigma}(z) 
- \epsilon_{k^{\prime}-\sigma} - \tilde{\Sigma}_{k^{\prime} -\sigma}(z)
+ \epsilon_{k^{\prime\prime}-\sigma} + 
\tilde{\Sigma}_{k^{\prime\prime} -\sigma}(z)
}
\ ,
\label{g0cij3}
\end{eqnarray}
\begin{eqnarray}
\chi(\epsilon_{k}, \epsilon_{k^{\prime}}, 
\epsilon_{k^{\prime\prime}}) =
(1-f(\epsilon_{k}))(1-f(\epsilon_{k^{\prime}}))
f(\epsilon_{k^{\prime\prime}})
+ f(\epsilon_{k})f(\epsilon_{k^{\prime}})
(1-f(\epsilon_{k^{\prime\prime}})) \ .
\label{chi1}
\end{eqnarray}
Here $\epsilon_{k\sigma} = \epsilon_{\sigma} + \epsilon_{k}$ is the
Hartree-Fock energy, $f(\omega)$ is the Fermi distribution function.

A way to simplify in Eq. (\ref{g0cij3}) the three-fold sum with 
respect to $k$ might be to introduce an approximate 
$\tilde{\Sigma}_{k\sigma}(z)$ whose $k$ dependence has been projected
onto the Hartree-Fock energy $\epsilon_{k\sigma}$, as follows.
\begin{eqnarray}
\tilde{\Sigma}_{\sigma}(\epsilon_{k\sigma}, z) =
\dfrac{
\int d\mbox{\boldmath$k$}^{\prime}
\delta(\epsilon_{k\sigma}-\epsilon_{k^{\prime}\sigma}) 
\tilde{\Sigma}_{k^{\prime}\sigma}(z)}
{\int d\mbox{\boldmath$k$}^{\prime}
\delta(\epsilon_{k\sigma}-\epsilon_{k^{\prime}\sigma})
} \ .
\label{peffk}
\end{eqnarray}
After the approximation 
$\tilde{\Sigma}_{k\sigma}(z) \approx 
\tilde{\Sigma}_{\sigma}(\epsilon_{k\sigma}, z)$,
Eq. (\ref{g0cij3}) is expressed as
\begin{eqnarray}
\overline{G}^{({\rm c})}_{0ij\sigma}(z) =
\int \dfrac{d\epsilon d\epsilon^{\prime} 
d\epsilon^{\prime\prime} 
\rho_{ij\sigma}(\epsilon)\rho_{ij-\sigma}(\epsilon^{\prime})
\rho_{ji-\sigma}(\epsilon^{\prime\prime})
\chi(\epsilon, \epsilon^{\prime},
\epsilon^{\prime\prime})}
{z - \epsilon - \tilde{\Sigma}_{\sigma}(\epsilon, z) - 
\epsilon^{\prime} - \tilde{\Sigma}_{-\sigma}(\epsilon^{\prime}, z)
+ \epsilon^{\prime\prime} + 
\tilde{\Sigma}_{-\sigma}(\epsilon^{\prime\prime}, z)}
\ .
\label{rpt01}
\end{eqnarray}
Here $\rho_{ij\sigma}(\epsilon)$ is the Hartree-Fock density of states
defined by $\rho_{ij\sigma}(\epsilon) = \sum_{k} \langle i|k \rangle 
\delta(\epsilon - \epsilon_{k\sigma}) \langle k|j \rangle$.

On the other hand, in the case of $\{ \lambda_{ij\sigma}=0 \}$ ({\it i.e.}, 
$\{ \overline{\lambda}_{ij\sigma}=1 \}$), 
it is not easy to obtain directly
a simplified expression of Eq. (\ref{g0ij2}) because
$\mbox{\boldmath$v$}_{\rm c}$ remains.  
However, $H^{({\rm c})}_{\rm I}(z)$ in Eq. (\ref{hci}) becomes the
Coulomb interaction in this case.
The second-order perturbation of
the temperature Green function yields then an approximate expression.  
\begin{eqnarray}
\overline{G}^{({\rm c})}_{0ij\sigma}(z) =
\int \dfrac{d\epsilon d\epsilon^{\prime} 
d\epsilon^{\prime\prime} 
\rho^{({\rm c})}_{ij\sigma}(\epsilon)
\rho^{({\rm c})}_{ij-\sigma}(\epsilon^{\prime})
\rho^{({\rm c})}_{ji-\sigma}(\epsilon^{\prime\prime})
\chi(\epsilon, \epsilon^{\prime},
\epsilon^{\prime\prime})}
{z - \epsilon - \epsilon^{\prime} + \epsilon^{\prime\prime}}
\ .
\label{rpt00}
\end{eqnarray}
Here $\rho^{({\rm c})}_{ij\sigma}(\epsilon)$ is 
the DOS for a cavity Green function for the Hamiltonian (\ref{tihc0})
with $\{ \overline{\lambda}_{ij\sigma}=1 \}$.

Therefore we obtain a simplified expression of the screened cluster memory
function, which is an interpolation between  Eq. (\ref{rpt01}) for
$\lambda_{ij\sigma}=1$ and Eq. (\ref{rpt00}) for $\lambda_{ij\sigma}=0$.
\begin{eqnarray}
\overline{G}^{({\rm c})}_{0ij\sigma}(z) = A_{ij\sigma}
\int \dfrac{d\epsilon d\epsilon^{\prime} 
d\epsilon^{\prime\prime} 
\rho^{({\rm c})}_{ij\sigma}(\lambda, \epsilon)
\rho^{({\rm c})}_{ij-\sigma}(\lambda, \epsilon^{\prime})
\rho^{({\rm c})}_{ji-\sigma}(\lambda, \epsilon^{\prime\prime})
\chi(\epsilon, \epsilon^{\prime},
\epsilon^{\prime\prime})}
{z - \epsilon - \lambda_{\sigma}\tilde{\Sigma}_{\sigma}(\epsilon, z) 
- \epsilon^{\prime} - 
\lambda_{-\sigma}\tilde{\Sigma}_{-\sigma}(\epsilon^{\prime}, z)
+ \epsilon^{\prime\prime} + 
\lambda_{-\sigma}\tilde{\Sigma}_{-\sigma}(\epsilon^{\prime\prime}, z)}
\ .
\label{lrpt01}
\end{eqnarray}
This is Eq. (\ref{lrpt0}).
The prefactor $A_{ij\sigma}$, the densities of states for
the cavity states $\rho^{({\rm c})}_{ij\sigma}(\lambda, \epsilon)$, 
and a simplified self-energy $\tilde{\Sigma}_{\sigma}(\epsilon, z)$
are given by Eqs. (\ref{aij}), (\ref{rhocav}),
and (\ref{peffk2}), respectively.
\vspace{5mm}

\end{document}